\documentclass{emulateapj}
 
\newcommand{\mycomment}[1]{}
\newcommand{\SDSSJ}{SDSS\,J1004+4112}
\newcommand{\degs}{\ifmmode ^{\circ}\else$^{\circ}$\fi}
\newcommand{\lopt}{\ifmmode L_{2500} \else $~L_{2500}$\fi}
\newcommand{\loglopt}{\ifmmode{\rm log}~L_{2500} \else log$~L_{2500}$\fi}
\newcommand{\logz}{\ifmmode{\rm log}~z \else log$~z$\fi}
\newcommand{\ax}{\ifmmode{\alpha_x} \else $\alpha_x$\fi} 
\newcommand{\aox}{\ifmmode{\alpha_{\rm ox}} \else $\alpha_{\rm ox}$\fi} 
\newcommand{\fcgs}{\ifmmode erg~cm^{-2}~s^{-1[B}\else erg~cm$^{-2}$~s$^{-1}$\fi}
\newcommand{\fnucgs}{\ifmmode {\rm erg~cm}^{-2}~{\rm s}^{-1}~Hz^{-1}\else erg~cm$^{-2}$~s$^{-1}$~Hz$^{-1}$\fi}
\newcommand{\lnucgs}{\ifmmode erg~s^{-1}~Hz^{-1}\else erg~s$^{-1}$~Hz$^{-1}$\fi}
\newcommand{\lcgs}{\ifmmode erg~~s^{-1}\else erg~s$^{-1}$\fi}
\newcommand{\kms}{\ifmmode~{\rm km~s}^{-1}\else ~km~s$^{-1}~$\fi}
\newcommand{\mone}{\ifmmode ^{-1}\else$^{-1}$\fi}
\newcommand{\mtwo}{\ifmmode ^{-2}\else$^{-2}$\fi}
\newcommand{\msun}{\ifmmode {M_{\odot}}\else${M_{\odot}}$\fi}
\newcommand{\lapprox }{{\lower0.8ex\hbox{$\buildrel <\over\sim$}}}
\newcommand{\gapprox }{{\lower0.8ex\hbox{$\buildrel >\over\sim$}}}
\newcommand{\nh}{\ifmmode{\rm N_{H}} \else N$_{H}$\fi}
\newcommand{\nhgal}{\ifmmode{ N_{H}^{Gal}} \else N$_{H}^{Gal}$\fi}
\newcommand{\nhintr}{\ifmmode{ N_{H}^{intr}} \else N$_{H}^{intr}$\fi}
\newcommand{\nhtot}{\ifmmode{ N_{H}^{tot}} \else N$_{H}^{tot}$\fi}
\newcommand{\atoms}{\ifmmode{\rm ~atoms~cm^{-2}} \else ~atoms cm$^{-2}$\fi}
\newcommand{\cmsq}{\ifmmode{\rm ~cm^{-2}} \else cm$^{-2}$\fi}

\newcommand\ha{\ifmmode {\rm H}\alpha \else H$\alpha$\fi}
\newcommand\hb{\ifmmode {\rm H}\beta \else H$\beta$\fi}
\newcommand{\oi}{\ifmmode{\rm [O\,II]} \else [O\,II]\fi}
\newcommand{\oii}{\ifmmode{\rm [O\,II]} \else [O\,II]\fi}
\newcommand{\ew}{\ifmmode{W_{\lambda}} \else $W_{\lambda}$\fi}

\shorttitle{LAMAS Reveals Small-Scale Absorbing Streamers in AGN} 
\shortauthors{Green et al.}

\received{2005 March 1}
\begin{document}


\title{Lens-Aided Multi-Angle Spectroscopy (LAMAS)
\\ Reveals Small-Scale Outflow Structure in Quasars} 


\author{Paul J. Green,}
\email{pgreen@cfa.harvard.edu}
\affil{Harvard-Smithsonian Center for Astrophysics, 60 Garden Street,
 Cambridge, MA 02138}



\begin{abstract}
Spectral differences between lensed quasar image components are
common.  Since lensing is intrinsically achromatic, these differences
are typically explained as the effect of either microlensing, or as
light path time delays sampling intrinsic quasar spectral
variability. Here we advance a novel third hypothesis: some spectral
differences are due to small line-of-sight differences through quasar
disk wind outflows.  In particular, we propose that variable spectral
differences seen only in component $A$ of the widest separation lens
SDSS~J1004+4112 are due to differential absorption along the
sightlines. The absorber properties required by this hypothesis are
akin to known broad absorption line (BAL) outflows but must have a
broader, smoother velocity profile.  We interpret the observed C\,IV
emission line variability as further evidence for spatial fine
structure transverse to the line of sight.  Since outflows are likely
to be rotating, such absorber fine structure can consistently explain
some of the UV and X-ray variability seen in AGN.  The implications
are many: (1) Spectroscopic differences in other lensed objects may be
due to this ``lens-aided multi-angle spectroscopy'' (LAMAS).  (2)
Outflows have fine structure on size scales of arcsec as seen from the
nucleus.  (3) Assuming either broad absorption line region sizes
proposed in recent wind models, or typically assumed continuum
emission region sizes, LAMAS and/or variability provide broadly
consistent absorber sizescale estimates of $\sim 10^{15}$\,cm.  (4)
Very broad smooth absorption may be ubiquitous in quasar spectra, even
when no obvious troughs are seen. 
\end{abstract}




\section{Introduction}
\label{intro}

\subsection{Quasar Outflows}
\label{outflows}

Evidence is accumulating that outflows occur wherever there is
accretion.  In active galactic nuclei (AGN) the outflows from
supermassive black holes (SMBH) are highly ionized, so their
signatures appear mostly in restframe ultraviolet (UV) and X-ray
spectra.  From spectroscopy of Seyfert nuclei, at least half show
narrow absorption lines (NALs) of highly ionized species in outflows
of $\sim$1000\kms.  From X-ray studies, warm absorber features are
present again in about half of quasar X-ray spectra 
\citep{Piconcelli05}, and many such features are seen in outflow
\citep{Crenshaw03a}. 

More massive, accelerating outflows create broad absorption lines
(BALs), whose P-Cygni profiles span velocities to $\sim$0.3$c$ and are
visible in the spectra of 15 -- 20\% of optically-selected quasars
\citep{Hewett03,Reichard03a} or more in unbiased samples
\citep{Chartas00}.  These outflows may exist in all quasars,
subtending a solid angle covering fraction at least as large
as their detection fraction.  

Studies of powerful mass outflows in quasars are rapidly reshaping our
understanding of physics near the supermassive black hole.  Sightlines
that pierce the absorbers yield information on the velocity and plasma
state of the gas that is impossible to obtain from emission lines,
most likely because the latter are formed by compounded emission from
much larger ensembles of clouds in a wide variety of physical states
\citep{Baldwin95}.   These broad line-emitting clouds lie at distances of
10$^{15} - 10^{18}$\,cm from the SMBH, spanning a range of densities
$n_e\sim$ 10$^8 - 10^{12}$\,cm$^{-3}$ and covering $\geq$10\% of the
ionizing source.  Recent models (e.g., \citealt{Elvis00}) hold
that the absorbing and emitting clouds are not only cospatial but
quite possibly identical, making knowledge of absorber properties even
more important to our understanding of AGN physics. 

For outflows, equatorial disk wind models (e.g., \citealt{Murray97,
Elvis00}) are currently favored, although
proponents of polar winds exist \citep{punsly05,hartnoll01}.  It has
often been suggested that  
outflows may contain a substantial kinetic luminosity (though see
\citealt{Blustin05}),  and impart significant mass and energy into the
interstellar (ISM) of the host galaxy and even into the surrounding
intergalactic medium (IGM; \citealt{Roy02}).   

A great deal more information about the structure and dynamics of the
outflowing absorber could be gleaned if some information about their
size scales were available.  However, the $\sim$parsec scale
emitting/absorbing regions near supermassive black holes (SMBHs) will
remain spatially unresolved for the foreseeable future (milliarcsec at
$z$$\sim$0.01).  If the dynamics of absorbers were known (e.g.,
Keplerian orbits in the SMBH potential), then spectroscopic
variability information might also lead to absorber size scales, {\em
  viz.} $a_V\sim v_{trans}\Delta t$.  Unfortunately, the absorber
dynamics are poorly constrained, and multi-epoch spectroscopy is 
difficult to arrange and therefore rare.  Gravitational lensing can
help.  

\subsection{Spectral Differences Between Lensed Images}

% SPECTRAL DIFFERENCES IN LENSED QUASARS
Because gravitational lensing is intrinsically achromatic, spectral
similarity is an early criterion to consider close quasar images as
lens candidates.  Redshifts, as well as emission and absorption line
profiles must be similar.  However, significant spectral differences
have been noted in {\em bona fide} lensed quasars (e.g., HE~2149-2745;
\citealt{burud02a}, SBS~1520+530; \citealt{burud02b};
\citealt{Oguri05}) with clearly identified lensing masses and some
with time delays (making the lens interpretation irrefutable).
Significant differences in absorber properties between image
components have been documented in optical/UV spectra of BALQSO lenses
(e.g., APM0829+5255; \citealt{Lewis02}, H1413+117;
\citealt{Angonin90}). \SDSSJ\, could represent a more extreme example
of this phenomenon due to its wider-angle separation, and/or to
fortuitous snapshots of unveiled phases.

One traditional explanation of such spectral differences is
microlensing, which preferentially magnifies parts of one image,
enhancing spectral components that originate from smaller emitting
regions (about the size of the Einstein radius of a star)
at the source.  Another possible explanation is intrinsic quasar 
spectral variability, combined with lensing time delays.

% SDSSJ1004
The recent discovery of the gravitationally lensed quadruple-image
$z=1.734$ quasar SDSS\,J1004+4112 \citep{Inada03,Oguri04} is exciting
because of its record-setting separation ( maximum of 14.6\arcsec\,
between images).  A $z$=0.68 cluster centered among the four lensed
images is confirmed as the massive lens.  More recently, deep ACS and
NICMOS images of \SDSSJ\, from {\em Hubble Space Telescope} ({\em
HST}) \citep{Inada05} have now revealed clear arcs, sheared images of
the quasar host galaxy, and a probable 5th quasar image, all of which
substantially constrain the plethora of viable lens models.  Here we
discuss primarily the 4 brighter image components $A - D$ with
existing spectroscopy.

In addition, there are intriguing differences between the spectra of
the 4 quasar images.   Both microlensing and variability have been
posited to explain the spectral differences, yet both explanations
have serious problems.   

Here we advance a novel third hypothesis that has only been mentioned
in passing (e.g., \citealt{Lewis98, Oguri04}) - line-of-sight 
differences through quasar outflows.  We propose that \SDSSJ\,
offers a revealing multi-angle view of quasar winds originating near
the nucleus, where absorption can change on small angular and time
scales.  While there are challenges also for this new hypothesis,
several of its predictions are immediately testable.  Given the
wide ramifications for AGN physics, it is well worth considering.  

In the next section, we briefly review the spectral evidence,
and troubles with the more traditional explanations. 
Then we examine the prospects and predictions of our own hypothesis.   

\section{Spectra of SDSS~J1004+4112}

\subsection{Description}

In SDSS~J1004+4112, the blue emission line wings of the brightest
image $A$ are enhanced relative to the spectra of image components
$B,C$ and $D$.  The ratios of component spectra reproduced in
Figure~\ref{ABCD}a from \citet{Oguri04} reveal that the differences
are larger for Ly$\alpha$$\lambda\,$1216/N\,V$\lambda\,$1240,
SiIV+OIV$]\lambda\,$1400 and C\,IV$\lambda\,$1549 lines than for
the lower ionization lines of CIII$]$$\lambda\,$1909 and
MgII$\lambda\,$2800. There are subtle 
differences between the ratio spectra $A/B$, $A/C$, and $A/D$, but the
general features are similar; the spectrum of $A$ is the most
divergent, with strong blue wings to the emission lines.  \footnote{We
note that, while not relevant to the present discussion,
\citet{Oguri04} discuss also the variable strengths of narrow
intervening e.g., MgII absorbers that can be seen in these spectra.}
The continuum flux ratios are approximately constant from 3000 -- 8000\AA.
  \footnote{Not all spectra were taken at the parallactic angle, so the
continuum slopes are suspect.  However, neither would we expect these 
parallactic angle effects to conspire to produce flat continuum
ratios.}  

Figure~\ref{ABCD}b, direct from from \citet{Richards04}, focuses on
multi-epoch spectra of the C\,IV emission line in components $A$ and
$B$.  The 7 spectra shown span observed-frame time delays of 322 days.
An enhancement of the blue wing of C\,IV in $A$ was seen in the first
(03 May 2003) spectrum which lasted at least 28d (since it is seen on
31 May), and then faded (since it was not there 21 Nov).  Not shown in
the figure are later epochs when the enhancement {\em reappeared} ---
seen in a spectrum of 26 Mar 2004, and again 10 and 19 April 2004
\citep{IAUC04, Wisotzki04}.

\subsection{The Microlensing Hypothesis}

Could the blue wing enhancement in $A$ be due to microlensing?
Microlensing of the broad emission line region (BELR) can occur
if it has structure smaller than the Einstein radius ($\sim3\times
10^{15}$\,cm for a 0.1$\msun$ star).  Several problems plague the
microlensing explanation: 
  [1] Microlensing should amplify not just the BELR, but also the
hot continuum-emitting region interior to it.  
While the single star approximation is poor for microlensing
at significant optical depth (the situation for multiply imaged
quasars), caustic models show a strong general correlation 
between the magnification of the continuum and of the BLR,
and it is extremely rare for the BLR to be magnified but not the
continuum \citep{Lewis04}. No amplification of the $A$ continuum was
seen ($<$20\%; \citealt{Richards04}) and the $A$ and $B$ continuua are
effectively identical \citep{Wisotzki04}.   
  [2] If the microlensing hypothesis is correct, then microlensing
does not act on the (smaller) continuum region but somehow acts only
on a select region of the BELR, and that this same configuration
{\em recurs}. The reappearance of the same enhancement renders
the microlensing explanation particularly unlikely.   
  [3] Strong line profile differences are also seen in the blue wings
of the lower ionization lines of CIII] and MgII \citep{Richards04}.
In most BELR models, these come from significantly larger regions than 
does C\,IV emission, and should be less susceptible to microlensing.
  [4] Line asymmetry induced by microlensing is expected for certain 
kinematic models of the BELR \citep{Abajas02}, the closest match being
for a modified Keplerian BELR with small lens size.  The $A$ profile
has varied strongly.  Similar profile variations would be expected in
other lensed systems but have never been seen.\footnote{While 
\SDSSJ\, is uniquely wide, this is irrelevant to microlensing.}   

We note that in the ACS and NICMOS images \citep{Inada05}, a dim
object (G4) is detected within $\sim$1\arcsec\, of image $A$ that 
could host a microlens.  However, it is uncertain whether or not the
object is a galaxy, and even more so whether it could provide a
microlensing optical depth at the position of image $A$ sufficient to
explain the observed variations. 

\subsection{Variability}

Spectral differences in many lenses are plausibly explained by
intrinsic variability combined with time delays, meaning that with
4 images, we are effectively viewing one quasar at 4 epochs.  Given
the asymmetry $(r_A-r_B)/(r_A+r_B)$ of the $A$ and $B$  images with
respect to  the lens, the {\em maximum} delay between them is
$\lapprox\, 30$ days \citep{Oguri04}.  However, $B$ never showed a blue
wing bump, although it persisted in $A$ for {\em at least} a month
\citep{Richards04}. 
Also, the persistently bluer profiles in $A$ should have appeared 
within $\sim$30d in $B$ (or disappeared in $A$ depending on which
image lags) but have not: the variability has been confined to $A$.
For these reasons, the usual hypothesis of intrinsic
variability plus lightpath time delay appears unlikely to be correct. 
However, the temporal coverage of spectroscopy to date can not
entirely rule out intrinsic variability plus time delay (see
\citealt{Richards04} for a full discussion). 

\subsection{Lens-Aided Multi-Angle Spectroscopy through Absorbing
  Fine Structure} 

Lensing geometry actually provides slightly different sightlines to
the lensed object, and the image separation $\theta$ is similar 
to the angular difference between the lines of sight 
as seen from the quasar nucleus $\Delta\phi$ \citep{Schneider92}.  
Indeed, \citet{Chelouche03} recently pointed out the utility of lensed
images of BALQSOs for probing absorber structure transverse to the
line of sight. 

We propose that small angular differences in sightline afforded by
lensing (lens-aided multi-angle spectroscopy; LAMAS hereafter) can
probe significantly different absorbing columns in 
quasars.  In the case of SDSS~J1004+4112, we propose that all 4 image
components suffer from absorption from a warm outflowing wind (similar
to that proposed by \citealt{Elvis00} and others), and that the
spectral differences are due to a line-of-sight to component $A$ that
pierces a persistently thinner and patchier part of the absorbing
flow.  A transient hole or ``rend'' in the flow along that sightline
is responsible for the blue wing C\,IV enhancement in $A$.

We note that a surprising Galactic analogy to LAMAS exists in the
massive young star $\eta$\,Carinae.  There is a dusty axisymmetric
bipolar ``Homunculus'',  formed by ejecta from $\eta$\,Car, which
creates a hollow reflection nebula.  Spatially-resolved {\em HST}
STIS spectra of the Homunculus
\citep{Smith03}, whose 3-dimensional structure is fairly well-modeled,
samples reflected light from the star emerging at different stellar   
latitudes and probing different parts of the outflow.  At viewing
angles separated on arc sec scales, large differences in the broad
($\sim$1000$\kms$) P~Cygni absorption profiles are seen.  Outflow
structure in a young massive star may well be qualitatively different
than in an AGN, but the dense line-driven wind and the
strong angular dependence of the absorption are analogous.
%   5 milli-arcsec or ~11\,AU at the distance of $\eta$\,Carinae. 

First we present evidence that the blue enhancements in \SDSSJ\, $A$
could be due to the alleviation of absorption along this sightline.
Second, we report on significant variations common in AGN absorbers.
Third, we address the geometry -- how different lines of sight may
probe significantly different absorber properties and what constraints
this and variability place on the absorber structure.  Fourth, we
address some possible objections to our hypothesis.

\section{Rending of the Veil}

\begin{figure}[h!]
% \plotfiddle{PSFILE}{VSIZE}{ROTANG}{HSCALE}{VSCALE}{HTRANS}{VTRANS}
%\plotfiddle{comparespec.eps}{3in}{0.}{400}{400}{-260}{30} 
%\plottwo{ABCDwratio.ps}{ABvary.eps}
\plotone{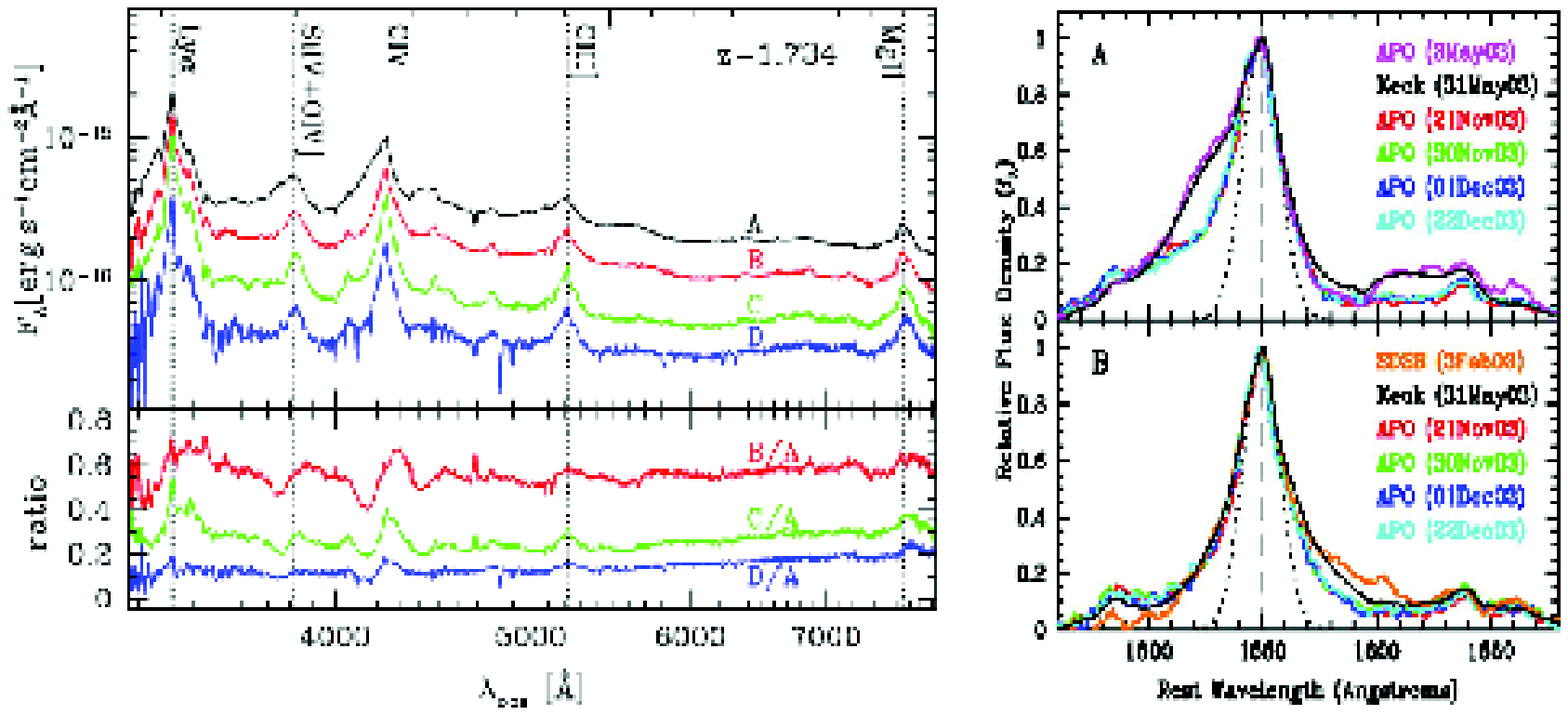}
%\vspace*{-1.25cm}
\caption{\small  {\bf SDSS~J1004+4112 SPECTRA} 
\underline{\em (a)} At left, spectra of the 4 quasar images of
SDSS~J1004+4112 rescaled for clarity (top) reproduced
from \citep{Oguri04}. The blue wings of emission lines are enhanced in
$A$, most notably in C\,IV. The bottom panel highlights the
differences, showing the ratio of each of $B$, $C$, and $D$ to the $A$
component spectrum.   
\underline{\em (b)} In the right panel, seven epochs of the C\,IV
emission line of SDSS~J1004+4112 are reproduced directly from
\citet{Richards04}.  The spectra 
have been smoothed, renormalized so their peaks match, and are
shown with a scaled Gaussian for reference (dotted line). 
The blue wing bump is apparent in the first 2 epochs of the $A$
spectrum. Even after 21 Nov 2003, $A$ maintains a strong {\em blue
asymmetry}.  The blue bump in component $A$ reappeared in 2004
\citep{IAUC04, Wisotzki04}.}
\vskip-0.2cm
\label{ABCD}
\end{figure}

\begin{figure}[ht!]
% \plotfiddle{PSFILE}{VSIZE}{ROTANG}{HSCALE}{VSCALE}{HTRANS}{VTRANS}
%\plotfiddle{balsratio2.eps}{6in}{0.}{500}{200}{200}{200} 
\plotone{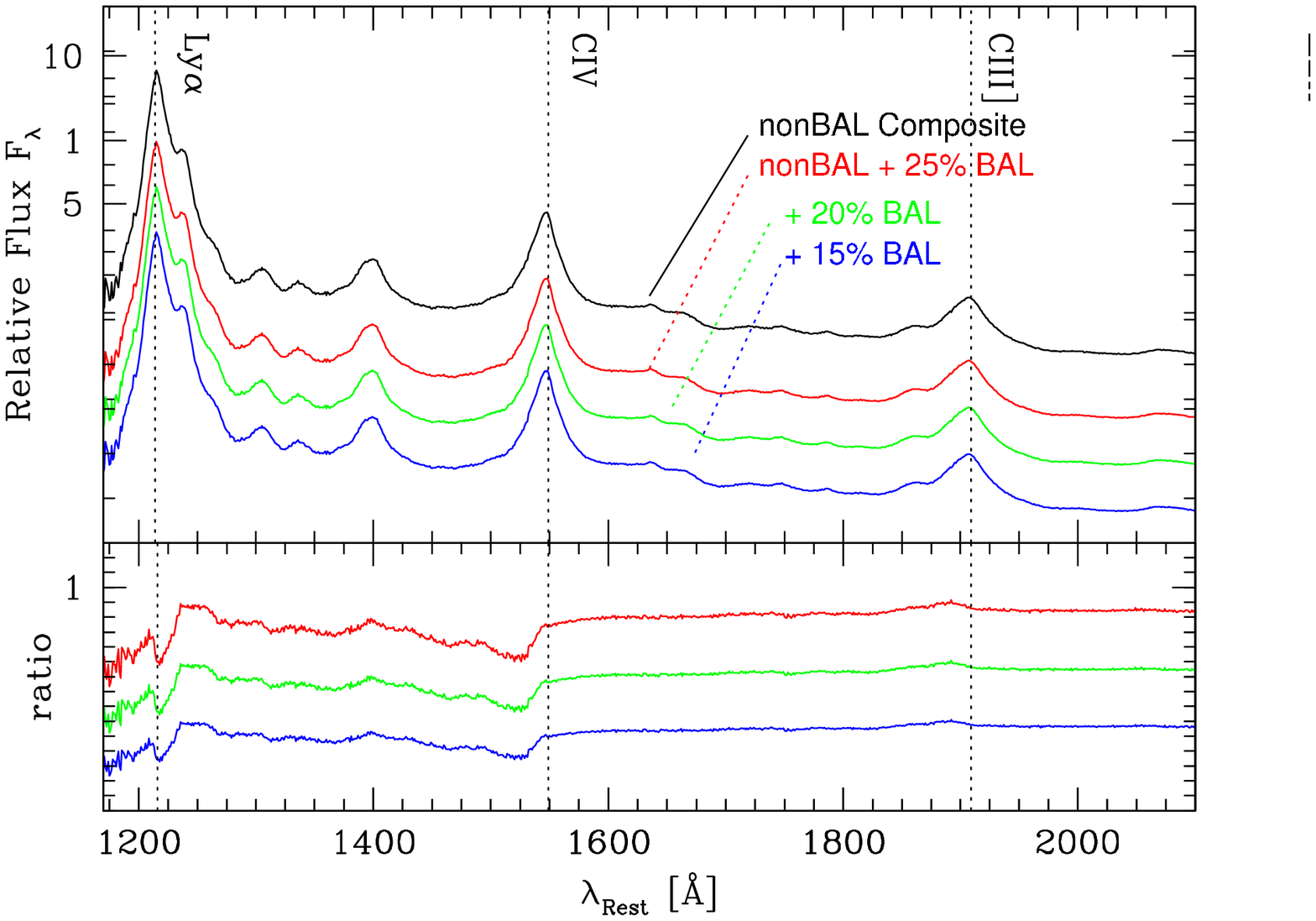}
%\vspace{-6cm}
\caption{\small TOP: A rest-frame non-BAL composite spectrum from
the SDSS from \citep{Reichard03b} (black line), along with composite
spectra we generated by adding 25, 20, and 15\% of the SDSS
high-ionization BAL composite spectrum (3 spectra below as marked).
We have taken the logarithm of flux and offset the spectra for clarity,
and we highlight just the region from Ly$\alpha$ to  CIII], where the
differences are most apparent.  No BAL troughs are evident even in the
25\% BAL spectrum. BOTTOM: The lower panel shows 
the ratio of these 3 part-BAL composite spectra to the non-BAL
composite.  While features may differ in detail, note the overall
similarity to the spectral ratios seen in Figure~\ref{ABCD} for
SDSS~J1004+4112. }    
\label{balcomp}
\end{figure}

The ratio spectra shown in Figure~\ref{ABCD}a are strongly reminiscent
of the structure seen in BALQSOs.  To illustrate this with typical
spectra, we take the the rest-frame non-BAL composite spectrum from
the SDSS \citep{Reichard03b}, assumed to be ``unabsorbed''.  Next we
generate ``part-BAL'' spectra by combining that composite with
the SDSS high-ionization BAL composite spectrum.  The non-BAL
spectrum is shown at the top of Figure~\ref{balcomp}, along with three
example part-BAL composite spectra generated by adding 25, 20, and
15\% of the SDSS high-ionization BAL composite spectrum.
Even though we plot the logarithm to make differences as visible as
possible, no clear absorption signature is evident even in the 25\%
part-BAL spectrum.  Then we divided the part-BAL composite by the
``unabsorbed'' (non-BAL) composite (lower panel of
Figure~\ref{balcomp}). While some features may differ in detail, note
the overall similarity to the spectral ratios seen in
Figure~\ref{ABCD} for SDSS~J1004+4112.  

The BAL and non-BAL composites have all their features smoothed and
broadened by the averaging process.  These composite ratio spectra (of
a part-BAL to a non-BAL composite) in Figure~\ref{balcomp} show strong
similarities to the \SDSSJ\, image component spectral ratios in
Figure~\ref{ABCD}.  This implies that absorption may account for the
differences along the sightlines to the different images.
Furthermore, there are 2 reasons to think that the absorbing outflows
in SDSSJ1004 may be smoother 
and broader than normally associated with BALs: (1) adding a
smooth/broad absorbed spectral component as in Figure~\ref{balcomp}
(TOP) does not create recognizable absorption signatures, nor are such
signatures visible in the $BCD$ spectra.  (2) The similarity of
the ratio spectra in Figures 1 and 2 suggest that the absorption troughs
in the $BCD$ components may also be smoothed and broadened,
as are those in the composites.  Normal BAL-type absorbers typically
show steep trough edges, which are clearly absent in \SDSSJ\,.
While the absorption in the BAL composite is smoothed by
averaging many sightlines to different normal BALQSOs, the absorber we
hypothesize to exist in \SDSSJ\, must be smoothed in velocity
for a different reason.  We propose that the absorber profile
is {\em intrinsically} smeared  broadly in velocity and consequently 
smooth onset and trailoff, i.e., with a profile more like a shallow
bowl than a trough.  Because such absorption is easily missed, this may be a
higher covering-factor but lower-column counterpart to more typically
remarked BAL flows. 

We propose that virtually all quasar sightlines penetrate a warm
(ionized), relativistically outflowing wind which is smooth in
velocity space but may be spatially clumped or filamentary.  These
ionized winds would quite normally sculpt the emission line profiles of all
quasars, particularly in the UV.  Excess emission in the blue wing
flux of resonance lines is not uncommon \citep{Bachev04}, and is
predicted in the models of \citet{Murray97} (see their Fig 7).  The
latter authors note that "observed profiles are even more
antisymmetric than the calculated profiles.  Our neglect of scattered
line photons may be partially responsible for this.  The absorption
troughs in BALs provide direct evidence for the existence of such
scattered line photons.''

In the case of SDSS~J1004+4112, our line of sight to $A$, as
influenced by the lensing geometry, probes through part of the
absorber whose column density is somewhat less dense, and also more
susceptible to the appearance of gaps in the patchy or perhaps
filamentary absorber.

\begin{figure}[ht!]
% \plotfiddle{PSFILE}{VSIZE}{ROTANG}{HSCALE}{VSCALE}{HTRANS}{VTRANS}
% \plotfiddle{2355m3357ra.eps}{6in}{270.}{500}{400}{00}{-500} 
\plotone{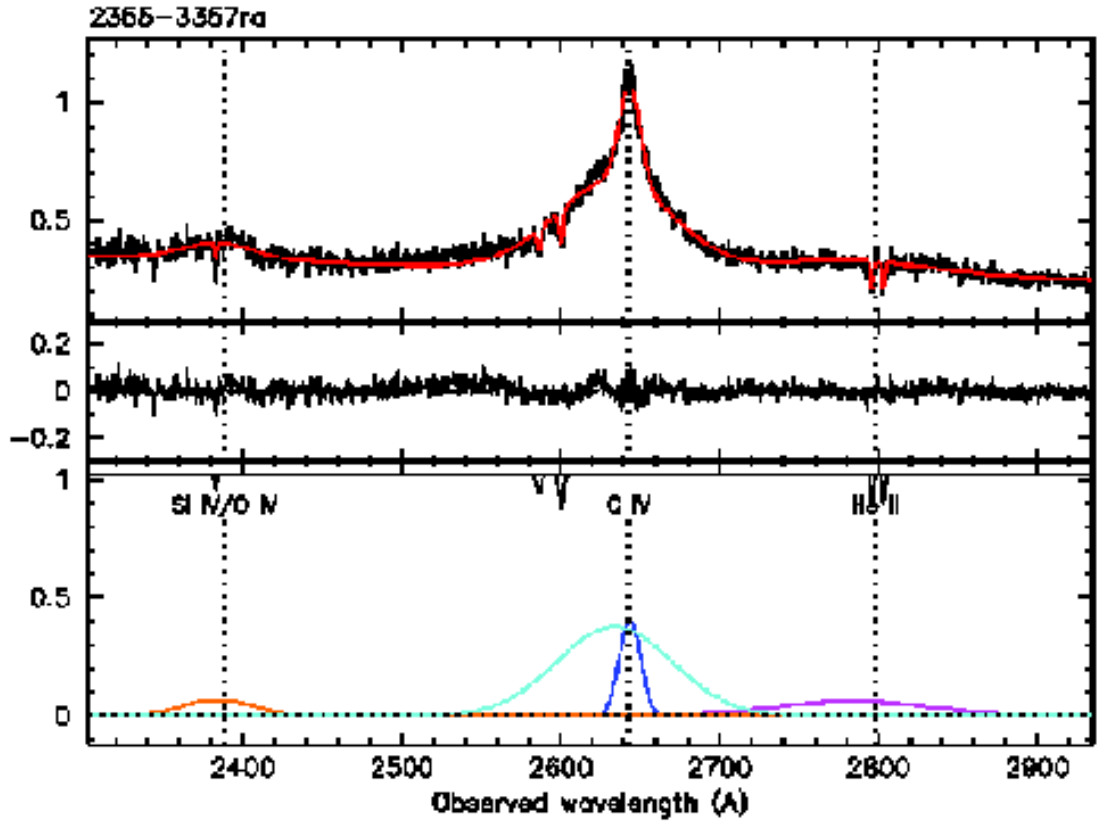}
\caption{\small  {\bf 2355--3357 } 
C\,IV region of the spectrum of the quasar PKS\,J2355-3357, with
multi-component fits from \citet{Kura02}.   The blue wings of C\,IV
emission line is enhanced here, with a profile similar to that in
\SDSSJ. } 
\label{2355m3357}
\end{figure}

As a microlensing effect, similar blue wing enhancements would not be
expected in unlensed quasars.  However as either a variability or a
line-of-sight effect, they should be seen (with as yet unknown
frequency) in the general (unlensed) quasar population.  We
investigated the emission line properties of AGN from the large sample
of {\em HST} FOS line measurements of Kuraszkiewicz et al. (2002,
2004).  While their spectral analysis was both careful and uniform,
the sample is heterogeneous, and our perusal subjective, so we cannot
draw firm conclusions on the fraction of objects with blue asymmetry
at this strength. However, we find several examples of C\,IV lines
with asymmetry similar the \SDSSJ\, spectrum seen in
Figure~\ref{ABCD}.  The spectrum and model fits to PKS\,J2355-3357 are
shown in Figure~\ref{2355m3357}.  This suggests that the same
phenomenon occurs in other objects along unlensed sightlines as
expected.  Because we have no multi-epoch spectroscopy of these
objects, we do not know whether the phenomenon is typically transitory.
%  A reasonable emission line fit is achieved using a blueshifted gaussian
%  for the broad line region component.

The absorbing structure we propose is not identical to a BAL flow as
typically observed: its velocity profile is broader and smoother.
This is because no steep trough edges are seen in Figure~\ref{ABCD}.
(Of course, no steep edges are seen in the comparison composite
spectrum of Figure~\ref{balcomp} either, but this is due to the fact
that the SDSS BAL composite is an average of 180 SDSS BALQSOs, whose
BAL troughs have a range of detachment velocities and velocity
profiles.)  Broad smooth outflow velocity profiles are plausible
because BALQSOs with unusually wide, smooth absorption have been
found.  Extreme examples like VPMS\,J1342+2840 \citep{Meusinger05},
SDSS\,J0105-0033 and SDSS\,J2204+0031 \citep{Hall02} lack any evidence
for the usual restframe UV emission lines, yet they show relatively
blue continua with no obvious dust reddening.  The most convincing
explanation to date of their spectral features is unusually wide,
overlapping low-ionization BAL troughs (loBALs).  Since these objects are quite
difficult to recognize and classify, they are likely severely
under-represented in existing AGN samples.  From our crude spectral
arithmetic in Figure~\ref{balcomp} and from the fact that most quasars
do not show X-ray absorption signatures as strong as in BALQSOs, the
flow we propose probably has lower overall column than typical BALs.
Indeed, there is evidence that the frequency of quasar absorption
increases towards lower columns \citep{Reichard03a}, making more
plausible a ubiquitous low-column smooth absorber.

The disk wind scenario \citep{Murray97, Elvis00} fits naturally into
this picture, since the absorbers are smoothly accelerating outflowing
sheets of warm ionized plasma. In the disk wind scenario, radiative
acceleration is in the radial direction, but the disk from which the
wind arises is rotating at approximately Keplerian speed until the
last marginally stable orbit \citep{Murray95}.  Put simply, the
outflows are almost certainly rotating, creating helical streamlines.
These are thought to be rising off the disk to create a ``martini
glass'' funnel-shaped outflow \citep{Elvis00}.

We might expect that holes or flow gaps would appear more commonly 
along the sightline to component $A$ of \SDSSJ\, if it is less
absorbed generally (see Figure~\ref{absmodel}).  Even after the
obvious blue wing bump subsided during later 2003 epochs shown in
Figure~\ref{ABCD}b, the blue/red flux ratios (measured
20\AA\, to either side of C\,IV$\lambda1549$) are $R_A=1.15$ and
$R_A=0.84$ (both $\pm 0.05$).  This supports our hypothesis of
persistently lower column along the sightline to the $A$ image
component. 

\subsection{Variable Absorption}

Time-dependent calculations \citep{Proga00} show that disk wind
instabilities result in a filamentary substructure to the flow, so we
naturally expect column density variations to occur along our
sightline, which would more strongly affect the blue line wings.
Rotation combined with clumpy outflows predicts absorber variability.
This prediction should be borne out in absorbers seen in unlensed
quasars.

In the UV, BALs have been seen to vary strongly in
QSOs.  At least one BALQSO spectroscopic monitoring campaign
has been performed \citep{Barlow94}, which showed
BAL variability in 15 of 23 BALQSOs.  Significant BAL variability
has been reported anecdotally elsewhere  (CSO203; \citealt{Barlow92},
SDSS\,0437-0045; \citealt{Hall02}).  The BALs in at least one quasar
have disappeared completely \citep{Ma02}. Variations may be due to a
combination of factors, including a change in line-of-sight column,
covering factor of the absorber, or ionization.   Variability
is well-documented and quite common in the narrower
absorbing systems seen in lower luminosity AGN (Seyferts) 
such as such as NGC3516 \citep{Koratkar96} and NGC4151 
\citep{Kraemer05} and others, as reviewed by \citet{Crenshaw03a}.

In the highly ionized circumnuclear environment, X-ray spectroscopy is
a sensitive probe of column density because metals in virtually any 
ionization state will absorb X-rays, whereas optical/UV
continuum absorption requires the presence of dust.  For instance, while the
continua of high-ionization BALQSOs are at most only mildly reddened
\citep{Reichard03b}, BALQSOs are strongly absorbed in X-rays
\citep{Green01,Gallagher02a}.  As measured by X-ray spectroscopy,
variations in the absorbing column density by factors of a few are
quite common in both optical broad and narrow line (Type\,I and II)
AGN.  In Seyfert~2 galaxies, 20 -- 80\% variability in the measured
absorbing column is endemic (23 of 24 objects; \citealt{Risaliti02}),
and is likely due to bulk motion of material across the line-of-sight.
\citet{Gallagher04} found a column variation of 6$\times
10^{22}$\atoms\, in BALQSO PG2112+059 over a $\sim$3yr timespan.
\citet{Gallagher02b} discovered hard-band variability at the 45\%
level on a timescale of 20\,ksec in the nearby mini-BAL QSO Mkn~231.
More dramatic changes are also seen.  UGC\,4203 underwent a transition
in its X-ray spectrum from Compton-thick to thin \citep{Guainazzi02},
and NGC\,3227 did the opposite \citep{Lamer03}. The authors suggested
the transition could be absorbing clouds or streams crossing our
line-of-sight.  An X-ray ``unveiling event'' was noted in NGC4388 by
\citep{Elvis04}, corresponding to a decrease in column of a factor of
100 in just 4 hours. Variability so rapid puts the NGC4388 absorber at
a few 100\,$R_S$ ($\sim 3\times 10^{15}$\,cm for
$M_{BH}=10^8\,M_{\odot}$), similar to the broad emission line region
or smaller. 

% Schwarzschild radius of a 1 solar mass black hole is 3e5 cm.
% Schwarzschild radius of a 1e8 solar mass black hole is 3e13 cm.
 
Absorber variability is clearly common if not endemic even in the
general (unlensed) AGN population.  Such variability should provide a
geometric constraint $a_V$ on the absorber sizescale from $\Delta\,t$
that is complementary to the LAMAS sizescale $a_L$ from $\theta$. 

\section{Geometry}

From the LAMAS perspective, the different absorber properties of the
two sightlines constrain the lateral sizescale $a$ of the
absorber as simply as $a_L \sim R_a\theta$, where $R_a$ is the
absorber distance and $\theta$ is the observed image splitting wherein
a signifcant absorbing column change is seen.  The difficulty here is
twofold.  First, there are at best only loose constraints on location
of the broad absorption line region (BALR): debate over $R_a$ still
spans 5 orders of magnitude from 0.01 to 1000\,pc
\citep{Elvis00,deKool01,Everett02}!  Second, the smooth, broad
absorbing flow we propose may span a different spatial regime than the
dense flows responsible for BALs (the BALR).

We first assume that $R_a\sim R_{BELR}$.  From reverberation mapping
\citep{Peterson97}, the size of the high-ionization (C\,IV) BELR in
NGC\,5548 is $R\sim$10\,lt-day, or $\sim 10^{16}$\,cm.  However, in
higher luminosity objects like \SDSSJ, reverberation time lags can
easily reach 100\,lt-day and more.  (The BELR size scales as $R_{BELR}
\propto 0.01\, L^{1/2}_{44}$pc from the central continuum source,
where $L_{44}$ is the 0.1 -- 1$\mu$m luminosity in units of
$10^{44}$\lcgs\, \citep{Netzer97, Peterson04}.)  If the absorber is at
a distance similar to the BELR (assuming 100\,lt-days), then the $A/B$
spectral differences seen in \SDSSJ\, across $\theta=3.7\arcsec$ imply
that cloud/stream columns can change significantly on size-scales of
$\sim 5\times 10^{12}$\,cm or $\leq 1$A.U. transverse to the line-of-sight.
% 100 lt-d = 2.592e+17cm
% 3.7/206265 = 1.79381e-05 rad
% a = R*Theta = 4.64955e+12 ; 1AU = 1.5e13cm

If we assume that the LAMAS and variability both probe the same
absorber size scale, how do these sizes compare?  Suppose the flow to
be in quasi-Keplerian rotation at $\sim 10^4$\kms\, (the median FWHM
of broad component of C\,IV emission; \citealt{Kura04}).  The observed
variability timescale of the C\,IV blue wing thus potentially presents
a {\em separate} ($\Delta\phi$) size constraint complementary to that
of LAMAS ($\theta$) discussed above: the blue wing bump disappearing
from $A$ in $<43$days (in the rest frame; \citealt{Richards04})
constrains the transverse size of a rend or flow gap to $a_V \lapprox
4\times 10^{15}$cm ($\sim$270\,AU).  If we again assume transverse
cloud motion created the variable line profile, but instead of
postulating a privileged sightline to $A$ we make the (unlikely)
assumption that with adequate monitoring, the same change {\em could}
have been seen to occur in components $B - D$, then the lensing delay
timescale is a more appropriate comparison.  Since most lensing models
for \SDSSJ\, in \citet{Oguri04} predict a time delay $\lapprox\, 30$
days, the result for $a_V$ is therefore similar.  Either way, $a_V$ is
inconsistent with (1000$\times$) the LAMAS sizescale $a_L$ based on
$R_a\sim R_{BELR}$ above.

% calc 30*24*3600*1e9 = 2.6e15 cm
% calc 2.592e+15/1.5e13 = 172.8 AU

% The crossing time of  a Keplerian cloud
% covering a region of size $r_{10}=r/10R_S$ around a black hole
% is $t_c \sim 0.13\,M^{1/2}_8\,d^{1/2}\,r_{10}$yrs
% \citep{Guainazzi02}, where $d$ is the cloud distance from the nucleus
% in pc. 

Using a single-phase wind to model the outflows from
FIRST\,J104459.6+365605, \citet{deKool01} found the low-ionization BAL
region must be at $R_a\sim$700\,pc.  In this case, the lateral
absorber size implied by LAMAS is $a_L\sim 4 \times 10^{16}$\,cm,
which is 10$\times$ larger than the variability sizescale.  
% 700*3.086e18 * 1.79381e-05 = 3.87499e+16 
However, the \citet{deKool01} sizescale is vary hard to reconcile with
partial covering seen in BALs.  \citet{Everett02} instead used a
multi-phase outflow model that successfully reproduced the many
absorption features (having different ionization parameters but
similar velocity structure) by placing the absorber at $\sim 4$pc from
the central source.  Using this value for $R_a$, LAMAS yields
($a_L\sim 4\times 10^{14}$cm or 30\,AU), about 10$\times$ {\em
smaller} than the variability estimate.

So if our LAMAS proposal holds true, the absorber distance is
inconsistent with most BELR distance estimates of $\sim 10^{16}$\,cm,
but falls somewhere between the best recent estimates of the absorber
distance $R_a$ for BALs of 5$\leq R_a \leq$700\,pc.

As an independent estimate, we consider that from an absorbed
sightline, the covering fraction of the absorber must be reasonably
large (between 0.1 and 1) for a small change in viewing angle to
significantly affect the absorption profile.  Put differently, the
size of the continuum region as seen from the absorber distance must
not be too large, otherwise a small change in viewing angle would not
detectably alter the absorption profile.  For \SDSSJ\, (with its
3.7\arcsec\, splitting), this implies that $R_a \geq 50000\, a_C$,
where $a_C$ is the projected UV continuum emitting region size.
Because $a_C$ is thought to be about $\sim 30-50\, R_S$, for an
accreting supermassive black hole of $10^8\, M_{\odot}$ we expect that
$a_C \sim 10^{15}$, yielding $R_a \geq\, 5\times 10^{19}$cm or
$\geq$10\,pc.  While independent from the above geometrical arguments,
this covering fraction argument under the LAMAS hypothesis
yields a similar plausible absorber sizescale (10$^{15}$\,cm).
Note that if spectra were to show the absorptionn of both
continuum and substantial BELR flux, then the absorber distance
implied by this latter argument would be quite large ($\gapprox
1$\,kpc). 

% 5e19*3.7/206265 = 8.96904e+14

\section{Objections} 

Several objections to LAMAS leap to mind.

\begin{figure}[h]
% \plotfiddle{PSFILE}{VSIZE}{ROTANG}{HSCALE}{VSCALE}{HTRANS}{VTRANS}
%\plotfiddle{absmodel.xfig.eps}{2in}{0.}{55}{55}{-190}{40} 
\plotone{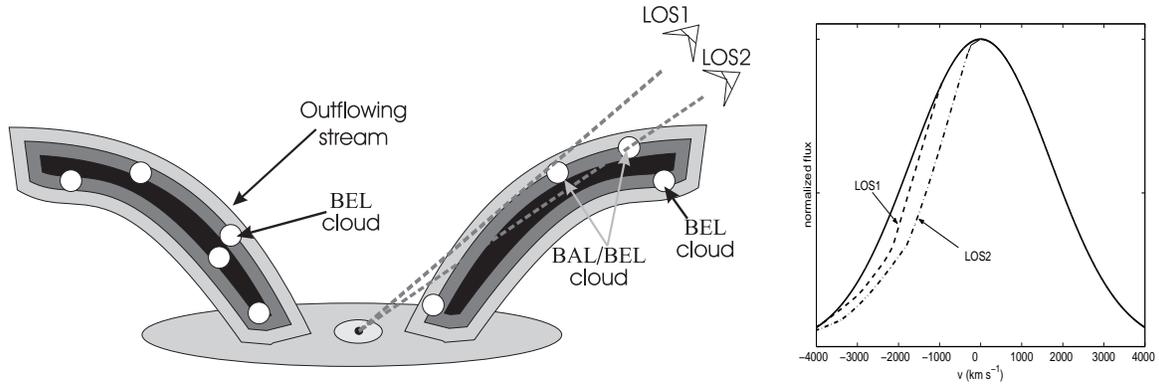}
%\vspace*{-1.5cm}
\caption{\small {\bf ABSORBER MODEL. } 
\underline{\em LEFT: ~} A cartoon of a stratified wind outflow model
\citep{Murray97, Elvis00, Everett02} shows 2 observers at slightly
different line-of-sight angles to the outflow.  In these models,
wind-embedded clouds/streams act as emitters (absorbers) when we look
at (through) them. \underline{\em RIGHT: ~} A Gaussian emission line
absorbed by a standard BAL outflow model (e.g., \citealt{Arav99}) along 2
slightly different sightlines as shown at left.}  
\vskip-0.5cm
\label{absmodel}
\end{figure}

{\em How could it be that the absorption changes on arcsecond sightline
scales but is identical in components $B - D$, some of which are more
widely separated from each other than from $A$?}  Our proposal is that
our sightline to $A$ is unique in this system, perhaps skirting the
edge of a larger-scale structure like a disk wind, as illustrated
crudely in Figure~\ref{absmodel}.   However, there is no reason in the
LAMAS picture that only one component could show a distinct absorption
profile. Spectroscopic monitoring of this and other lenses may well
identify such cases, which would help confirm the model.  Our
absorption interpretation would be most sensitively verified
by simultaneous UV and X-ray monitoring of spatially-resolved image
components in this and other lenses. 

{\em \SDSSJ\, is not a BALQSO.}  Most or possibly
{\em all} QSOs contain BAL-type outflows \citep{Hamann93}.  But the
classic deep BAL troughs are observed only for sightlines traversing
dense BAL streams \citep{Ogle99}, whereas lower-column parts of the
flow cover more solid angle and may affect the emission lines less
spectacularly in {\em most} QSOs \citep{Green98}. 
\citet{Reichard03a} noted that the fraction of quasars with BALs
increases strongly towards lower BALnicity, so that "the fraction of
quasars with intrinsic outflows may be significantly underestimated."
Broader, smoother outflows such as proposed here be even harder 
to detect, but could be ubiquitous. 

{\em If such smooth, broad outflows are ubiquitous, what special
conditions yield more typical BALs, which are detached blueward of 
the emission line peak velocity? } Since their measured X-ray column
densities are highest of all types of QSOs, the densest part of the
wind is probed by BAL-type outflows.  In a disk wind structure al\`a 
\citet{Elvis00}, the arch of the BAL wind (as it moves from
predominantly vertical to radial velocity) means that it is already
accelerated when it first crosses our sightline
(Figure~\ref{absmodel}).  This need not be true for the WHIM that
transports the BAL material.  The WHIM can pervade a much larger
opening angle, and be accelerated from a lower velocity. 

{\em If smooth, broad outflows are ubiquitous, why aren't
most quasars absorbed by columns as large as $\sim${\rm 10}$^{22}$
in the X-ray regime?}  X-ray spectral fitting to absorption {\em
features} in bright optical- and radio-selected quasars  typically
yields measured columns of $\sim$10$^{21}$ \citep{Reeves00}.  However,
very broad features from warm ionized gas would not be easily detected,
just as they are unrecognized in the majority of UV spectra.  In fact,
for Seyfert 1s and radio-quiet quasars lacking obvious strong
absorption signatures in the UV and X-ray regimes, there is a
ubiquitous and poorly explained broad soft spectral excess below about
2\,keV.  \citet{Gierlinski04} and \citet{Sobolewska05} have
interpreted the soft excess as an artifact of previously
unrecognized, relativistically smeared, partially ionized absorption;
strong very 
broad O\,VII, O\,VIII, and Fe absorption features at 0.7 -- 0.9\,keV
can lead to an apparent upward curvature below these energies,
mimicking soft excess emission.  Whether such flows are related is
still unclear.  Their proposed X-ray absorber 
velocities ($v/c\sim 0.2$) greatly exceed what we see in the UV spectra
of SDSSJ1004  ($v/c\sim 0.04$), but the absorbing zones may be
disjoint \citep{Crenshaw03b} or stratified \citep{Steenbr05}, or the
velocity range may vary between objects. 

{\em Some significant details of the 
part-BAL composite spectra in Figure~\ref{balcomp} 
 differ from those of \SDSSJ .}
Figure~\ref{balcomp} is meant to be illustrative only,
because there are several reasons why those spectra differ in some of
the details:  1) The BAL composites are rather heavily smoothed
(by dint of averaging many BALs with different velocity profiles).  Weaker
features are thereby blended away, and then further diluted by the
weighting in the part-BAL composite. 2) Details of the composites
depend on their method of construction: how the mean is weighted, in
what rest wavelength region the contributing spectra are normalized,
the dereddening algorithm, and how the quasars' systemic velocity is
defined for deredshifting. Furthermore, the SDSS sample selection and
spectroscopy means that  QSOs contributing to the composite have
varying luminosity, redshift, and intrinsic reddening as a function of
rest wavelength \citep{Willott05}. 3) The SDSS composites match SDSSJ1004 in
neither luminosity nor redshift, so that related emission line differences
are expected (e.g., the Baldwin effect \citealt{Yip04,Green96})
in CIII] and HeII.  4) True BALQSOs typically show other differences
in emission line profiles and strengths relative to non-BALQSOs that
are possibly related to an enhanced accretion rate (e.g.,
\citealt{Boroson02}). 5) The LAMAS outflow may sample different
regions than typical BAL flows, perhaps with different launch point
and/or acceleration profile.  Assuming a (e.g., biconical) disk-wind
outflow model, our sightline to \SDSSJ\,$A$ may be somewhat farther
from the denser part of the outflow, especially given the claim of
repeated 'rending' of the flow.  In this part of the outflow (e.g.,
the lower column warm highly ionized medium - the 'WHIM') the 
pressure/ionization balance may be different than in a typical BAL
flow.  Given so many caveats, the accordance between the BAL
composites and Figure~\ref{ABCD} seems rather remarkable.  And
certainly no plausible, testable combination of emission line
production and microlensing theory is yet available that can
reproduce the \SDSSJ\, spectra as well.

\section{Caveats}

Whether due to different effective orientations (caused
by lensing-induced sightline difference) or to time variability,
differential absorption by itself may not be able to produce three
key features seen in the spectral differences between components.
First is the apparent absorption in CIII$]$$\lambda\,$1909 and
He\,II$\lambda$\,1640. Second, the LAMAS model implies that emission
lines in most quasars would have an intrinsic blueward asymmetry
except for typical broad smooth absorption caused by warm outflowing
winds. If true, emission-line profiles {\em not} subject to resonant
absorption (e.g., CIII$]$$\lambda\,$1909) should retain that
asymmetry in the general AGN population. Finally, we discuss
the apparent observed change in the red wings of the emission-line
profiles (between images and also temporally).  

1) 
The blue wing of the C\,III] emission profile also appears to be
enhanced in image A relative to the other images (see Figure 4 of
\citealt{Richards04}), but there should be no absorption (and so no
differential absorption) detectable from the semi-forbidden C\,III]
transition.  However, this spectral difference may still be due to
absorption from another species, since BALQSOs can show some
spectral signatures there.  This is evident for instance in the
plots like Fig~10 of \citet{Richards02}, where their BALQSO
composites show some significant C\,III] line strength and profile
differences from the non-BAL composite.  The trough seen near
restframe 1909\AA\, in \citet{Richards04} Fig 4 is not necessarily
from CIII].  FeIII absorption (UV34,UV48) are strong iron lines in
the CIII] region \citep{Graham96}. Such lines have been seen in
absorption in BALQSOs 
e.g., Fig\,22 of \citet{Hall02}, though those tend to be in the more
rare low-ionization BALQSOs (loBAL QSOs) only.\footnote{We note that
spectral differences are seen in Mg\,II in Fig~\ref{ABCD}a, and
Mg\,II absorption is the criterion for a loBAL designation.}  For
unknown reasons, strong Fe\,II and Fe\,III {\em emission} blends are
endemic to BALQSOs generally, and correlate with BAL strength
\citep{Weymann91}.\footnote{Such lines are also seen in emission in
other types of AGN, but are most easily detected when narrow, for
example in Narrow Line Seyfert 1s like IZwI \citep{Laor97} and
quasars like 2226-3905 \citep{Graham96}. }  So, some differences in
iron blend profiles are not unexpected if the $BCD$ component
spectra are more BALQSO-like.
 
Also troubling is that the \SDSSJ\, ratio spectra appear to 
show an absorption feature in HeII$\lambda\,1640$ which is 
rare in BALs.  Some effect {\em related to} absorption may be 
at work here.  For instance, for reasons as yet unknown, the He\,II
emission feature in BALs, and in objects with larger CIV blueshifts is
weaker and broader in general (c.f. BALQSOs and composite C in
Figures~4 and 10 of \citealt{Richards02}).  

2) 
In the general AGN population, emission-line profiles that are {\em
not} subject to resonant absorption (e.g., CIII$]$$\lambda\,$1909)
should retain the blueward asymmetry implied by the differential
absorption hypothesis.  While this is a cogent objection to the
differential absorption hypothesis, it has only weak empirical
evidence available to test it because there are very few strong
non-resonant emission lines seen in AGN spectra.  CIII$]$ is
the strongest, and even in the absence of strong iron blends, it is
most often heavily blended on the blue side by SiIII$]$$\lambda~$1892
and Al\,III$\lambda~$1862 (e.g., \citealt{Vest01}).  The best resolved
example is probably Ark~564 \citep{Leighly04}, which indeed shows
isolated symmetric CIII$]$ lines at the systemic redshift, which 
mates poorly with the differential absorption hypothesis.  However,
this very narrow line Seyfert~1 is hardly representative itself.

3) 
The observed excess in the red wings of the emission-line
ratios in Fig~\ref{ABCD}a is significant, and nothing similar
is seen in Fig~\ref{balcomp}.  At least for CIV, the emission peak is
often blueshifted in BALQSOs: features in composite spectra of quasars
with larger blueshifts correlate with composites of increasing
BAL-type (hi towards loBAL) absorption (Fig.10 in
\citealt{Richards02}.  This works against a hypothesis of differential 
absorption, because $B,\,C,\,D$ as the more absorbed components should
show blueshifted, not redshifted emission profiles.

\section{Implications and Predictions}

An appealing aspect of our model is that it is subject
to immediate observational tests. X-ray observations of BAL
samples \citep{Green01,Gallagher02a}, measure intrinsic absorbing columns
of $\nhintr \geq 10^{22}$\cmsq, much higher than naively 
derived from the (typically saturated) UV BALs.  
Since we propose that $A$ is seen via a typically less-absorbed
sightline, $A$ could show a less absorbed X-ray spectrum, and also  
larger $f_X/f_{opt}$ relative to components $B-D$.  
Differential absorption studies of
lenses have been difficult in X-rays, due to the close
($\sim1\arcsec$) spacing of lensed images \citep{Morgan01, Chartas02}, but  
are quite feasible with wide lenses like \SDSSJ.  An 80\,ksec X-ray
observation of \SDSSJ\, was performed by Chandra in January 2005.
The one caveat to this prediction is that the highly-ionized,
high velocity flow proposed may require high $S/N$ X-ray spectra
to detect and model correctly \citep{Gierlinski04}.  

Since we further propose that $A$ is susceptible to show gaps in the
absorber flow, X-ray measurements during a similar UV blue emission
line asymmetry event could show substantially lower absorption.  A
very crude estimate based on Figure~\ref{balcomp}, and typical BAL
columns as measured in the X-rays of $\leq$10$^{23}$ \citep{Green01}
suggests a column change of $\sim$10$^{22}$.  Alternatively, X-ray
spectral fits might also reveal a change in {\em covering factor} of
the absorption. 

Wide quasar pairs at similar redshifts are under intense study as
possible wide lenses and have been hunted in large surveys like the
2dF \citep{Miller04} and SDSS (\citealt{Inada03,Oguri05}).  In a large enough
statistical sample of lenses, LAMAS predicts that the
degree of spectral difference should correlate with separation angle,
and should be independent of proximity to a bright galaxy of
high microlensing optical depth.  The same can be said
for the variability + time delay interpretation, but in that
case, adequate spectroscopic monitoring should reveal 
propagation of spectroscopic features to all image components.

Perhaps the most controversial part of this proposal is its
implication that {\em all} quasar spectra might be self-absorbed at
some level by their smoothly outflowing winds, but that this fact has
escaped our notice since quasars were discovered some 40 years ago
\citep{Schmidt63}.  However, the origin of observed
emission line profiles and even the quasi-power-law continuum of
quasars remain in general poorly understood.  Both
their overall similarity and their diversity are difficult to
explain \citep{Baldwin95}.  Absorbing winds may be a crucial missing
component. 

Variable absorption may also be endemic.  The salient characteristics
of quasar variability seem to be (1) timescales of months to years (2)
lower luminosity quasars have larger amplitudes of variability (3) at
all redshifts, variability is larger at shorter wavelengths (quasars
are bluer when brighter) (4) variability amplitudes increase with
redshift, indicating evolution of the quasar population or the
variability mechanism \citep{deVries05}.  To explain (non-blazar)
variability, models include accretion disk instabilities (e.g.,
\citealt{Kawaguchi98}), so-called Poissonian processes, such as
multiple supernovae (e.g., \citealt{Terlevich92}), and gravitational
microlensing (e.g., \citealt{Hawkins93}).  Since absorption has been
seen to vary significantly on timescales of about a month, then some
fraction of quasar flux variability should be due to absorbers.
However, this does not require that more absorbed QSOs are more
variable.  Indeed, BALQSOs show no different variability properties
than QSOs generally \citep{vandenberk04}. Although the absorbers
discussed here are warm (ionized), dust may also be expected to
contribute to varying degrees which would contribute to the reddening
of dimming events.  According to \citet{Elvis02}, quasar outflows are
natural dust producers.  Since quasars are expected to be dustier at
early cosmological times \citep{Wilman00}, this may naturally explain
the correlation of variability with redshift.

\section{Conclusions}

We propose that small shifts in line-of-sight to quasars
afforded by gravitational lensing can yield noticeable differences
in spectroscopic line profiles, due to lateral fine structure in
quasar absorbing outflows. 
Our hypothesis that lens-aided multi-angle spectroscopy
of \SDSSJ\, probes a rotating, generally smooth, sheet-like flow
with small-scale lateral spatial structure is consistent with these
facts: 
  (1) spectroscopy of lensed BALQSOs generally shows similar
BAL trough profiles, but exceptions are known
  (2) significant absorber column differences can exist between  
sightlines to lensed BALs
  (3) absorption columns to individual quasars are known to vary
significantly on timescales of hours to years
  (4) one component of \SDSSJ\, shows a UV emission
line blue wing enhancement that is spectroscopically consistent with
an unveiling event 

Models of multi-epoch spectroscopic observations of lensed quasar
images are necessarily convolved with models of intrinsic quasar
variability, and with the as-yet poorly constrained structural
parameters of the emitting/absorbing regions.  Microlensing holds some
promise for resolving the structure of the BELR in quasars.
Unfortunately, detailed and accurate lens models are required.  Worse,
the size, mass, and velocity of the microlensing caustic will remain
poorly constrained because microlensing depends on events (caustic
crossings) that are fundamentally unpredictable and irreproducible,
requiring a statistical approach to deconvolve the BELR structure. In
contrast, if the LAMAS dominates over microlensing, then only the 
easily-measured angular separation affects the observations. 

The most common proposed causes of spectroscopic differences in lens
components (time variability and microlensing) are both proven
phenomena, while LAMAS is not.  LAMAS is distinguishable in several
ways from these, and we propose several tests above.   
The addition of LAMAS to the mix may seem to muddy the waters, 
but LAMAS should be taken seriously until disproved, and may in fact
provide direct geometric information about the internal structure of
AGN that has long eluded us. 

The author gratefully acknowledges support through NASA contract
NAS8-03060 (CXC).  Many thanks to Tom Aldcroft, Doron Chelouche, 
Adam Frank, and Josh Winn for illuminating conversations and
collegiality.  Thanks also to the anonymous referee who
was extremely thorough and highlighted issues with the 
LAMAS model that will help test its viability in future 
observations and modelling.  

{}

\clearpage

\clearpage

\clearpage

\clearpage

\end{document}